\documentclass[twocolumn, times, showpacs,preprintnumbers,amsmath,amssymb,prl]{revtex4}

\usepackage{graphicx}%{epsfig}% Include figure files

\usepackage{dcolumn}% Align table columns on decimal point

\usepackage{bm}% bold math

\usepackage{times}% bold math

\usepackage{color}
\usepackage{colortbl}

\begin{document}

\title{Sub-Doppler Laser Cooling of Thulium Atoms in a Magneto-optical Trap}

\author{D.~Sukachev*}
\author{A.~Sokolov}
\author{K.~Chebakov}
\author{A.~Akimov*}
\author{S.~Kanorsky}
\author{N.~Kolachevsky*}
\author{V.~Sorokin*}
 \affiliation{P.N. Lebedev Physical Institute, Leninsky prospekt 53, 119991 Moscow, Russia\\*Moscow Institute of Physics and Technology,
141704 Dolgoprudny, Moscow reg. Russia}

\begin{abstract}{
We have experimentally studied sub-Doppler laser cooling  in a magneto-optical trap for thulium atoms working at the wavelength of 410.6\,nm. Without any dedicated molasses period of sub-Doppler cooling, the cloud of $3\times 10^6$ atoms at the temperature of 25(5)\,$\mu$K was observed. The measured temperature is significantly lower than the Doppler limit of 240$\mu$K for the cooling transition at 410.6\,nm.  High efficiency of the sub-Doppler cooling process is due to a near-degeneracy of the Land\'e-$g$ factors of  the lower $4f^{13}6s^{2}\, (J\,=\,{7}/{2})$ and the upper $4f^{12}5d_{3/2}6s^{2}\, (J\,=\,{9}/{2})$ cooling levels.}
\end{abstract}

\maketitle

{\bf 1. Introduction.}
Laser cooling of atoms is an important step towards solution of a number of fundamental and applied problems, such as quantum condensation~\cite{Ketterle_BEC}, study of cold atoms interactions~\cite{Lett0, Weiner}, atomic interferometry~\cite{Peters}, as well as development of microwave and optical atomic clocks~\cite{Clairon, Oates}.  In a number of applications, atoms should be confined in relatively shallow magnetic~\cite{Migdall} or optical~\cite{Grimm, Katori} traps. In many cases the depth of such traps does not exceed 1$\mu$K and for their efficient loading the initial temperature of an atomic cloud should be of the order of 10$\mu$K. Similar initial temperatures are called for in atomic interferometers and atomic fountains to minimize ballistic radial spread of an atomic cloud~\cite{Rile}.

The sub-Doppler laser cooling is commonly used for loading such dipole traps because it provides the highest phase-space density at the minimal depth of a dipole trap~\cite{DipoleTrap}. In most experiments, a dedicated procedure of sub-Doppler cooling is implemented. This process takes place directly after loading atoms in a magneto-optical trap (MOT), and starts with switching off the quadruple magnetic field of the MOT, after which the red frequency detuning from the cooling transition is gradually increased while the light intensity is reduced.

\begin{figure}[h]
\centerline{
\includegraphics[width=0.45\textwidth]{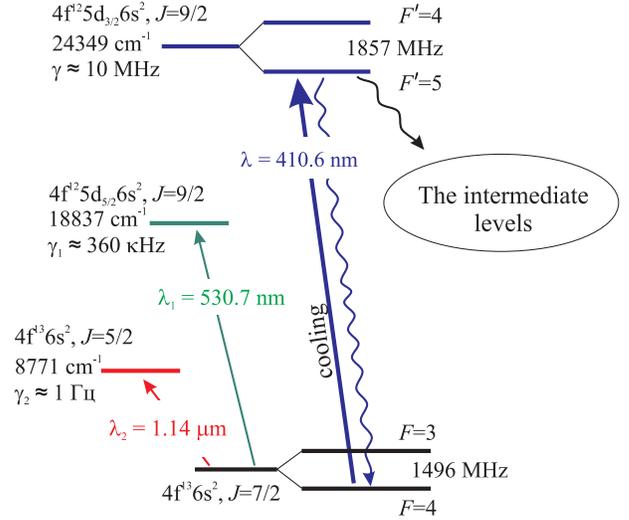}}
 \caption{Fig.1. Tm energy levels. $\lambda$, $\lambda_1$, $\lambda_2$ are transitions wavelengths and $\gamma$, $\gamma_1$, $\gamma_2$ are natural linewidths of the transitions.}
 \label{levels}
\end{figure}

Formerly, we have demonstrated laser cooling of thulium atoms in a MOT~\cite{Sukachev}. Lanthanides with the hollow $4f$-electronic shell possess a number of  intriguing properties like big dipole magnetic moments of the ground state as well as the presence of the shielded ground state fine structure sublevels coupled with magneto-dipole optical transitions (e.g. the transition at 1.14$\mu$m in Tm shown in Fig.\,\ref{levels}).

In this Letter the Doppler (Section 2) and the sub-Doppler (Section 3) laser cooling mechanisms are briefly reviewed. In Section 3 the influence of a magnetic filed on the sub-Doppler cooling efficiency is discussed, and in Section 4 we discuss the experimental results obtained on sub-Doppler laser cooling of Tm atoms {\it directly in a MOT} without any dedicated sub-Doppler cooling procedure. \\[1pt]

{\bf 2. Doppler cooling.}
The equilibrium temperature of the cooling process is determined by a balance of the cooling and heating processes caused by absorption and reemission of laser light. Originally, the Doppler theory of laser cooling was elaborated~\cite{Letochov} which considered a two-level atom. In case the light intensity is less than the saturation intensity of the cooling transition, the final temperature of atoms has the following dependence on the frequency detuning of cooling light from the resonance $-\delta$:
\begin{equation}
T(\delta)=\frac{h\gamma}{2k_B}\frac{\delta^2+\gamma^2/4}{\gamma\delta},
\label{DopplerZeroLimit}
\end{equation}
where $\gamma$ denotes the natural linewidth of the cooling transition and $k_B$  is the Boltzman constant. This temperature has a minimum at $\delta = \gamma /2$ which is referred to as the Doppler limit:
\begin{equation}
T_D={h\gamma}/{2 k_B}.
\label{DopplerLimit}
\end{equation}
For instance,  the Doppler limit  for the resonance cooling transition in cesium (which is widely used for laser cooling) is of 120$\mu$K.  Such temperatures proved to be too high for applications listed in the Introduction. It is necessary to use additional methods to decrease a temperature like a technique called sub-Doppler laser cooling.\\[1pt]

{\bf 3. Sub-Doppler cooling.}
Pioneering experiments with sodium atoms in optical molasses\cite{Lett1} showed that laser cooled atoms possessed lower temperatures than predicted by the Doppler theory. For atoms  possessing a complex structure of ground state magnetic sublevels, another cooling mechanisms take place. These additional mechanisms increase the cooling rate compared to the Doppler cooling one, which results in much lower temperatures\cite{Dalibard}. The lowest temperature which can be reached by the sub-Doppler cooling approaches {\it the recoil limit} $T_{\rm rec}=h^2/2\lambda^2mk_B$, which is much lower than the Doppler limit ($m$ is atomic mass). For instance, the recoil limit for cesium is 100\,nK.

Development of sub-Doppler cooling techniques open the possibility to use laser-cooled atoms in a number of fundamental and applied tasks (see the Introduction).

For the sub-Doppler cooling on the fine structure transitions like $F \to F+1$ ($F$ is total atomic momentum) the resulting temperature has the following dependency~\cite{Minogin}:
\begin{equation}
T\propto \frac{I}{F\delta}.
\label{eq:MinoginTemperature}
\end{equation}

Unlike the temperature dependency in the Doppler theory (\ref{DopplerZeroLimit}), which has the  minimum (\ref{DopplerLimit}) at the detuning  $\delta = \gamma / 2$, the sub-Doppler temperature dependency (\ref{eq:MinoginTemperature}) monotonically decreases with the detuning  $\delta$. One can reliably determine whether the sub-Doppler cooling takes place by measuring the temperature dependency. The answer will be unambiguous even in the presence of systematic errors which may  lower the temperature.

Sub-Doppler cooling mechanism is very sensitive to magnetic fields~\cite{Valentin, Shang, Walhout}. Without magnetic fields the cooling forces of  the Doppler and the sub-Doppler cooling mechanisms work together, both of them become zero for atoms with zero velocity. In the presence of magnetic fields, the Doppler cooling force is equal to zero for the so-called {\it locking} velocity of:
\begin{equation}
v_D = -g_e \frac{\mu_B B }{\hbar k},
\label{DopplerVelocity}
\end{equation}
While the locking velocity for the sub-Doppler mechanism is:
\begin{equation}
v_S = -g_g \frac{ \mu_B B }{\hbar k},
\label{SubdiopplerVelocity}
\end{equation}
where $g_e$  and $g_g$  stand for the Land\'e $g$-factors of the upper and the lower cooling levels, $\mu_B$ denotes the Bohr magneton and $k$  is the wavenumber. It should be noticed that the sub-Doppler cooling process is efficient only for the tiny velocity range near $v_S$ . If the velocities $v_D$ and $v_S$  differ much, only a small number of atoms on the wing of Maxwell-Boltzmann distribution undergoes the sub-Doppler cooling, almost without influencing the total temperature of atoms. The difference between $v_D$ and $v_S$ grows with the increasing of an external  magnetic field which results in disabling the sub-Doppler cooling process at the magnetic field on the order of 1\,G~\cite{MagneticInhibition}.

This effect impedes reaching sub-Doppler temperatures directly in the MOT since the atomic cloud resides not at the zero of the MOT quadrupole magnetic field due to its finite diameter and non-ideal alignment of the MOT laser beams. Field gradients in the MOT are usually about 20\,G/cm, so the displacement of 1\,mm is enough to block the sub-Doppler cooling process. In this case most experiments use an additional specific sub-Doppler cooling stage.

Efficient sub-Doppler cooling directly in the MOT is possible if the Land\'e $g$-factors of the upper and the lower cooling levels are nearly equal~\cite{Er}.
In our case the relative difference of the Land\'e $g$-factors of the cooling levels in Tm is only 2\% which enables efficient sub-Doppler cooling.

Besides the sub-Doppler mechanisms, there exist other methods to reduce temperature below the Doppler limit. For example, secondary Doppler cooling on weak transitions~\cite{Katori1} facilitates decreasing temperature of atoms with degenerative magnetic structure of the ground state ($^{20}$Mg, $^{40}$Ca, $^{88}$Sr). However, it requires additional stabilized laser source tuned to the corresponding weak atomic transition.\\[1pt]

{\bf 4. Experiment.}
We use experimental setup described in Ref.\cite{Sukachev}. Thulium atoms were loaded in the MOT from an atomic beam preliminary decelerated in a Zeeman slower~\cite{K.Chebakov}. We used the classical MOT scheme with three orthogonal couples of counter-propagating laser beams with $\sigma^-$ and $\sigma^+$ polarizations. The cooling transition is $4f^{13}6s^{2}\, (J\,=\,{7}/{2},\, F\,=\,4)\to 4f^{12}5d_{3/2}6s^{2}\, (J\,=\,{9}/{2},\, F\,=\,5)$ at 410.6 nm. For the cooling light we use the second harmonic of Ti:Sa laser. The frequency detuning of cooling light could be varied in the range of several $\gamma$ around the transition, where $\gamma$ = 10(4) MHz is its natural linewidth. Compared to Ref.\cite{Sukachev}, we replaced a photomultiplier tube by a CCD-camera for detection. The atomic cloud was projected on the CCD with the magnification of 1:1. The observation axis was in the horizontal plane at the angle of 45$^{\circ}$ to the horizontal cooling beams. CCD was triggered by a signal microcontroller driving  the whole cooling process.

The  CCD-camera enabled us to carefully adjust the MOT. We obtained symmetrical atomic cloud of Gaussian profile with the radius  r = 80 $\mu$m  (at the $1/e$ level).
The cloud typically contained $3 \times 10^6$ atoms, which corresponds to a density at the MOT center of $10^{12}$\,cm$^{-3}$.

\begin{figure}[t!]
\includegraphics[width=0.45\textwidth]{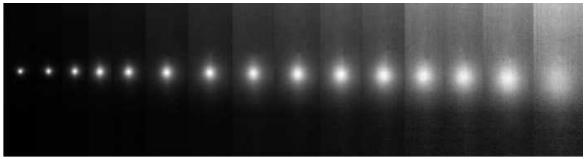}
 \caption{
 A number of successive shots of expanding cloud after switching off laser beams and magnetic fields.
 The photos captured after the time intervals $\Delta t$ = 0, 0.5, 1, 1.5, 2, 2.5, 3, 3.5, 4, 4.5, 5, 5.5, 6, 7 и 8\,ms.}
 \label{expanding}
\end{figure}

The MOT temperature was measured by expansion of the atomic cloud after switching off all magnetic and light fields. After some  time interval $\Delta t$ of the ballistic expansion the cloud was illuminated by a short (200 $\mu$s) probe laser pulse tuned  in the resonance with the cooling transition. The typical intensity of the probe beam was 100\,mW/cm$^2$ . The MOT images taken after different ballistic expansion time are shown  in Fig.\,\ref{expanding}.

\begin{figure}[h]
\centerline{
\includegraphics[width=0.45\textwidth]{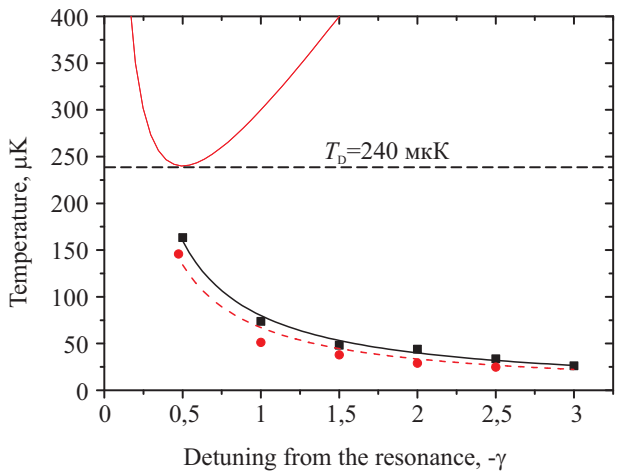}}
 \caption{
The dependence of atoms temperature on laser beams detuning at two values of the saturation parameter $S = I / I_{sat}$, where $I_{sat} = 18$\,мВт/см$^2$ is the saturation intensity. The squares correspond to $S = 2$ and the circles correspond to  $S = 0.4$. Dashed lines are the theoretical model (\ref{eq:MinoginTemperature}). The upper curve represents the temperature in the Doppler theory (\ref{DopplerZeroLimit}).}
 \label{fig1}
\end{figure}

For the ballistic expansion, the cloud radius $r_x(t)$(at the level $1/e$) changes in time as:
\begin{equation}
r_{x}(t)=\sqrt{r_{x}(0)^{2}+\frac{2k_B T}{m}\times t^{2}},
\label{eq:sigma_T}
\end{equation}
where $T$ is the atomic temperature. Measuring the time dependence of cloud radius we calculated the initial temperature using (\ref{eq:sigma_T}). The dependency of the atomic temperature on the laser frequency detuning $\delta$ is shown in Fig.\,\ref{fig1}.

From Fig.\,\ref{fig1}, the temperature is gradually decreases with increasing the red detuning. It proves that efficient sub-Doppler cooling takes place in the MOT. The lowest measured temperature was 25(5)\,$\mu$K.

\begin{figure}[h]
\centerline{
\includegraphics[width=0.45\textwidth]{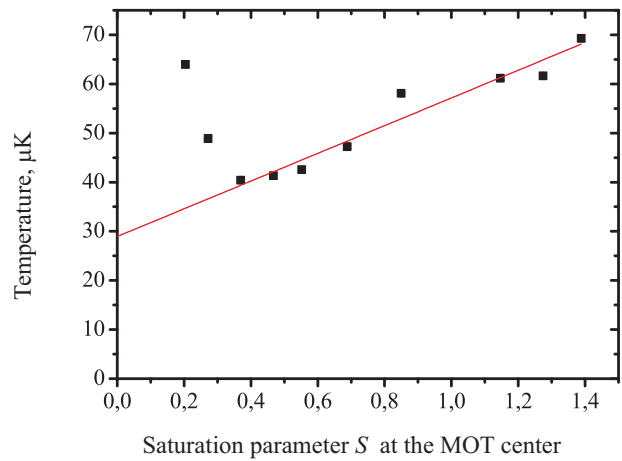}}
\caption{
The dependence of atoms temperature on the total light intensity at the trap center. The frequency detuning of the cooling beams is $-\gamma$. The straight line is the theoretical model (\ref{eq:MinoginTemperature}).}
\label{fig3}
\end{figure}

Temperature dependency on the intensity of the cooling beams is shown in Fig.\,\ref{fig3}.
Higher temperatures observed at low saturation parameters may result from unstable behavior of the MOT in this regime. At higher intensities the temperature linearly grows according to (\ref{eq:MinoginTemperature}). The linear extrapolation to zero intensity yields the nonzero temperature which indicates the presence of some additional heating process caused, perhaps, by non-ideal degeneracy of the Land\'e $g$-factors $g_e$ and $g_g$.\\[1pt]

{\bf 5. Conclusion.}
We have experimentally investigated  the sub-Doppler laser cooling process in a magneto-optical trap for thulium atoms. It is shown that due to the specific level structure of thulium atom, the  sub-Doppler cooling mechanism efficiently works directly in the MOT without the additional cooling cycle. The lowest observed temperature was 25(5)\,$\mu$K for the cloud containing $3 \times 10^6$  atoms and having the radius of 80\,$\mu$m. The corresponding phase-space density is of $10^{-5}$.

These results show, that efficient sub-Doppler cooling will also take place on the weak cooling transition $4f^{13}6s^{2}\, J\,=\,{7}/{2} \to  4f^{12}5d_{5/2}6s^{2}\, J\,=\,{9}/{2}$ at 530.7\,nm.
For this transition the relative difference of Land\'e $g$-factors is only of 0.8\%. Second-stage laser cooling on this transition will result in temperatures down to microkelvin scale (the Doppler limit is 8\,$\mu$K) and will increase the MOT lifetime. In our current MOT the lifetime is about 1\,s which is limited by unavoidable population leaks from the upper cooling level to the neighboring levels of opposite parity~\cite{Sukachev}. The magneto-optical trap at 530.7\,nm will not have this drawback.

Moreover, such low temperatures facilitate loading atoms in optical dipole trap and investigating atom interactions at low velocities.

This work is supported by President Grant (MD-3825.22009.2), RFBR (grand 09-02-00649a) and Program of fundamental research of Presidium RAS "Extreme light fields and their applications".

\end{document}